\documentclass[aps,prl,twocolumn,10pt,superscriptaddress,longbibliography]{revtex4-2}
\usepackage[utf8]{inputenc}
\usepackage{textcomp}
\usepackage{amsmath}
\usepackage{amssymb}
\usepackage{graphicx}
\usepackage{physics}
\usepackage{stackrel}
\usepackage{multirow}
\usepackage{braket, mleftright}
\usepackage{empheq}
\usepackage{subfigure}
\usepackage{soul}
\usepackage{blkarray}
\usepackage{bm}
\usepackage{bbold}
\usepackage{color}
\usepackage{algorithm}
\usepackage[normalem]{ulem}
\usepackage{algpseudocode}
\usepackage[unicode=true,
 bookmarks=false,
 breaklinks=false,pdfborder={0 0 1},backref=false,colorlinks=false]
 {hyperref} 
\usepackage{dsfont}
\usepackage{upgreek}

\makeatletter

\usepackage{amsfonts}
\usepackage{amsthm}
\usepackage{graphics}
\usepackage{bm}
\usepackage{color}
\usepackage{latexsym}
\usepackage{subfigure}
\usepackage{bbm}
\usepackage{xcolor}
\usepackage{xparse}
\usepackage[framemethod=TikZ]{mdframed}

\usepackage{indentfirst}
\usepackage{tabularx}

\newcommand{\quotes}[1]{``#1''}

\newcommand{\rket}[1]{| #1 )}
\newcommand{\rbra}[1]{( #1 |}
\newcommand{\rbraket}[2]{( #1 \, | \, #2 )}
\NewDocumentCommand{\rketbra}{O{#2} m}{{| #1 )}{( #2 |}}

\newcommand{\id}{\mathbbm{1}}
\newcommand{\eg}{\emph{e.g.}~}
\newcommand{\ie}{\emph{i.e.}~}

\makeatother

\definecolor{ingo}{rgb}{.8,.5,0}

\newcommand{\TII}{\affiliation{Quantum Research Center, Technology Innovation Institute, Abu Dhabi, UAE}}
\newcommand{\UFRJ}{\affiliation{Instituto de F\'{i}sica, Universidade Federal do Rio de Janeiro, P.O. Box 68528, Rio de Janeiro, Rio de Janeiro 21941-972, Brazil}}

\begin{document}

\title{Robust ultra-shallow shadows}

\author{Renato M. S. Farias}
\email{renato.msf@gmail.com}
\TII
\UFRJ

\author{Raghavendra D. Peddinti}
\TII

\author{Ingo Roth}
\TII

\author{Leandro Aolita}
\TII

\begin{abstract}
We present a robust shadow estimation protocol for wide classes of low-depth measurement circuits that mitigates noise as long as the effective measurement map including noise is locally unitarily invariant. This is in practice an excellent approximation, encompassing for instance the case of ideal single-qubit Clifford gates composing the first circuit layer of an otherwise arbitrary circuit architecture and even non-Markovian, gate-dependent noise in the rest of the circuit. 
We argue that for weakly-correlated local noise, the measurement channel has an efficient matrix-product representation, and show how to estimate this directly from experimental data using tensor-network tools, 
eliminating the need for analytical or numeric calculations. 
We illustrate the relevance of our method with both numerics and proof-of-principle experiments on an \textrm{IBM Quantum} device. 
Numerically, we show that unmitigated shallow shadows with noisy circuits become more biased as the depth increases. 
In contrast, using the same number of samples, robust ultra-shallow shadows become more precise with increasing depth for relevant parameter regimes.
The gain in sample efficiency is still limited by the noise per gate, resulting in an optimal circuit depth per noise level.
Experimentally, we observe improved precision in two simple fidelity estimation tasks using five-qubit circuits with up to two layers of entangling gates, by about an order of magnitude.
Under the practical constraints of current and near-term noisy quantum devices, our method maximally realizes the potential of shadow estimation with global rotations and identifies its fundamental limitations in the presence of noise.
\end{abstract}

\maketitle

\paragraph{Introduction.}
\label{sec:introduction}
Randomized measurements are a central tool for the estimation of observables on repeatedly prepared quantum states \cite{Eisert2020, Roth2021, Elben2022}. 
Applications abound \cite{Brydges2019, Huggins2022, Huang2022, Sack2022, Berg2022, Seif2023, Jnane2023, Abbas2023, Rath2023, Vermersch2023, Vitale2023, Denzler2023, Zhou2020, Liu2022}. 
Ref.~\cite{Huang2020} derived guarantees and proved optimality of linear estimation from bases measurements 
generated by global or local Clifford rotations and coined the term \quotes{classical shadows}. 
Besides optimality, an important contribution of Ref.~\cite{Huang2020} is the efficiency of estimation from global Clifford rotations. 
For example, with the same classical shadow from constantly many state copies in the system size $n$, one can estimate many different pure state fidelities.
In practice, however, harnessing this efficiency is typically not possible, since global Clifford rotations require the accurate implementation of \quotes{long} circuits of $\order{n^{2} / \log(n)}$ many local gates \cite{Aaronson2004}.
This obstacle has motivated the study of \quotes{shallow shadows}, where shorter circuits are employed to \quotes{interpolate} between local and global rotations \cite{Hu2023, Bu2022, Akhtar2023, Arienzo2023, Bertoni2024}.
There is evidence that for certain observables 
classical shadows with global rotations can be approximately de-randomized to $\mathcal{O}(n\log(n))$ gates \cite{Bertoni2024}. 
On noisy intermediate-scale quantum (NISQ) hardware, however, even measurement circuits in this scaling regime are challenging.  
For the measurement, one typically wants to work with hardware-efficient circuits of very low depth. For clarity, we refer to such construction as \emph{ultra-shallow shadows}. 

The two-design property of uniformly random Clifford rotations allows for straight-forward analytical inversion of the effective measurement map \cite{Scott2006},
a prerequisite to constructing the classical shadow. However, the works on shallow shadows resort to numerics for the evaluation of the effective measurement map for depth larger than two \cite{Arienzo2023}, exploiting the local invariance of the circuit ensembles \cite{Hu2023, Bu2022, Akhtar2023, Bertoni2024}. 
However, for measurement circuits involving entangling gates, noise, and (calibration) errors will typically cause sizable biases in the derived estimators on current NISQ hardware. 
For this reason, using ultra-shallow circuits for precise estimations in practice unavoidably requires a variant of classical shadows that mitigates circuit noise. 
That such mitigation is possible for classical shadows with single-qubit or global random Clifford rotations if the noise is well-characterized was shown for simple depolarizing noise models \cite{Koh2022}, local gate-dependent noise models \cite{Brieger2023}, and via randomized benchmarking experiments \cite{Onorati2024} 
\footnote{Note that related but differently motivated research lines are mitigating noise in the state preparation using non-linear shadow estimation \cite{Huggins2021, Hu2022, Seif2023, McNulty2023, Zhou2024, Liu2024, Grier2024}, or existing error-mitigation techniques \cite{Jnane2024}, and the optimization of dual frames \cite{Fischer2024}, e.g.}.
Going beyond these works, ideally one wishes for a robust shadow scheme that replaces analytical or numerical calculations of the measurement map with a direct empirical estimation of it in the presence of noise.
Such a tool was developed in Ref.~\cite{Chen2021}, again for the cases of single-qubit or global random Clifford rotations under the assumptions of gate-independent Markovian noise acting after the rotation. 
However, a careful extension to the most crucial, experimentally relevant regime of ultra-shallow circuits, where no analytical calculation of the measurement frame is, in general, available, is missing. 
Despite the large body of literature on classical shadows, a central question is still open: 
Can the theoretically established improvement in sample efficiency of randomizing measurements with entangling gates be turned into a practical advantage in near-term experiments?

Our work answers this question in the affirmative. We show that achievable advantages are significant but also limited at current noise levels. 
To this end, we present a robust shadow estimation scheme for ultra-shallow circuits with arbitrary architectures of random local gates, under mild and practically plausible assumptions on the noise. 
We assume only that the average noisy measurement channel is invariant under local rotations and the accurate implementation of a single initial state.  
A simple but experimentally crucial observation is that, for this to happen, it is sufficient for instance that the first circuit layer consists of accurately implemented single-qubit random Cliffords, regardless of the subsequent circuit, which can even be subject to time-stationary, gate-dependent, non-Markovian noise. 
As observed in previous works, single-qubit Pauli invariance guarantees that the measurement map is Pauli diagonal and the circuit shallowness guarantees that the corresponding diagonal elements admit a tensor train (TT) description of low-rank. 
For us, it is crucial that this structure is maintained if the noise is approximately local.
This allows us to develop efficient methods for directly estimating the measurement channel using standard tensor-network tools such as the modified alternating least squares ($\operatorname{MALS}$) \cite{Oseledets2012}.  

We numerically quantify how unmitigated shadow estimations (of both fidelities and Pauli observables) in the presence of noise become increasingly biased as the circuit depth increases. 
In contrast, we show that, for experimentally relevant regimes, robust ultra-shallow shadows become more precise as the depth increases up to an \emph{optimal depth} depending on the strength of the incoherent noise.  
For instance, for depth $D = 2$ and $n=7$, we observe that with our method average errors of fidelity estimation with random target states improve by a factor of two over noise-free local shadows.
Without mitigation, errors of shallow shadows are about an order of magnitude larger than with our method. 
We find that, even for these small depths and system sizes, infidelities of the entangling gates of at most $10^{-3}$ are required to realize improvements in precision at all. 
Moreover, we apply our scheme to a recent experiment \cite{Huggins2022} where ideal, four-qubit global Clifford rotations were approximated by noisy circuits of depth five, in an unmitigated fidelity estimation. 
In numerical simulations, we observe a four-fold reduction of error with our method. 
Finally, we perform a proof-of-principle experiment on the \texttt{ibm\_nairobi} device. 
We estimate the effective measurement map and find that, indeed, simplified assumptions beyond what our methods exploit are not permissible.

Our framework offers a versatile and practical solution to classical shadows in the most relevant regime for near-term quantum devices, where randomized measurements are implemented by noisy shallow circuits. 

\paragraph{Robust classical shadows.}
\label{sec:robust_classical_shadows}

We start by reviewing the classical shadow protocol on an ideal noiseless device. 
Consider a quantum device that can be repeatedly prepared in an unknown $n$-qubit quantum state $\varrho$.
Furthermore, we will work with a quantum device that is capable of implementing circuits composed from a unitary gate set $\mathcal{G}$ and of measuring in a fixed (computational) basis, specified by the effects $\mathcal B = \{\Pi_{z} = \ketbra{z} \}_{z \in [d]}$, with $d = 2^{n}$ and $[d] = \{0, 1, \dots, d - 1\}$.

A probability measure $\nu$ on the unitary group $\mathrm U(d)$ with support in the implementable circuits defines an ensemble of random bases measurements. 
The corresponding \emph{measurement protocol on an input state $\varrho$} repeats the following primitive: 
($i$) draw $g \sim \nu$; 
($ii$) prepare $\varrho$, apply $g$, and perform the measurement $\mathcal B$; 
($iii$) observing measurement outcome $z$, record the tuple $(z,g)$. 
The output of the data acquisition is a sequence $\mathcal S = \{(z_k, g_k)\}_{k=1, \hdots |\mathcal S|}$, with $|\mathcal{S}|$ the cardinality of $\mathcal{S}$.

Every iteration of the protocol is described by the same positive operator-valued measure (POVM) taking values $\Pi_{z,g} = g^\dagger \, \Pi_{z} \, g$ with $g$ in the support of $\nu$ and $z \in [d]$. 
If the support of $\nu$ is sufficiently large, the POVM is informationally complete (IC), and asymptotically any information about $\varrho$ can be reconstructed from the measurement statistic $\mathcal{S}$.

To be instructive, in the following, we restrict ourselves to the simplest task of estimating a linear functional $(O | \varrho) = \Tr[O^{\dagger} \varrho]$, 
here defined as the Hilbert-Schmidt inner product with an endomorphism $O$ of $\mathbb{C}^d$.
Examples include the estimation of the fidelity of $\varrho$ with respect to (w.r.t.) a pure target state $O = \ketbra{\psi}$ or multi-qubit Pauli observables.
The robust shallow shadows explored in our work can also be readily employed in non-linear estimation (see \eg Ref. \cite{Garcia2021}). 

Given $\mathcal{S}$, an estimator for $\rbraket{O}{\varrho}$ can be constructed via standard linear inversion \cite{Waldron2018}:
Consider the frame operator of the measurement POVM, also known as the effective measurement map,
\begin{align}
  S = \mathbb E_{g \sim \nu}  \left[\sum_{z\in [d]} \rketbra {\Pi_{z,g}} \right]\,,
\end{align}
where we have used Dirac notation w.r.t.\ the Hilbert-Schmidt inner product. 
Since the POVM is informationally complete (IC), $S$ is invertible and the elements of the canonical dual frame are given by $\bar{\Pi}_{z,g} = S^{-1}(\Pi_{z,g})$, 
fulfilling $\mathbb E_{g\sim \nu} {\big[}\sum_{z \in [d]} \rket {\bar\Pi_{z,g}} \rbra{\Pi_{z,g}}  {\big]}= S^{-1} S = \id_{d^{2}}$, with $\id_{d^{2}}$ being the $d^{2} \times d^{2}$ identity operator.
In the classical post-processing, we evaluate the scalar function $o(z,g) = (O | \bar\Pi_{z,g})$ on each entry of $\mathcal S$ and take the empirical mean $\hat{o} = \sum_{(z,g) \in \mathcal S} o(z,g)$. 
For the ideal device, we observe $(z,g)$ with probability density $p(z,g) = (\Pi_{z,g} | \varrho) \, \mathrm{d} \nu(g)$ and, thus, $\hat{o}$ is indeed an unbiased estimator of 
\begin{align}
  \mathbb{E}_{z,g}(\hat{o}) = \mathbb{E}_{g\sim \nu} \Bigg[ \sum_{z \in [d]} \, (O | \bar\Pi_{z,g})(\Pi_{z,g} | \varrho) \Bigg] = (O | \varrho) \, .
  \label{eq:equation_unbiased}
\end{align}

In practice, however, experimental imperfections in the implementation of the basis measurement $\mathcal B$ and the circuits in the support of $\nu$ induce a bias in the estimator. 
We denote by $\tilde{\Pi}_{z,g}$ the elements of the POVM that are implemented on the device and define the \quotes{one-sided} \emph{noisy frame operator} as 
\begin{align}
  \tilde{S} = \mathbb E_{g \sim \nu} \left[\sum_{z\in [d]} | \Pi_{z,g})( {\tilde\Pi_{z,g}}|\right] \, .
  \label{eq:noisy_frame_definition}
\end{align}
The same argument above now yields a non-vanishing estimator bias such that $\operatorname{bias}(O,\varrho) = |\mathbb{E}(\hat{o}) - (O | \varrho)| = |(O | S^{-1} \tilde{S} - \id_{d^{2}}|\varrho)|$.

The form of the bias already hints at a conceptually simple solution to mitigate the noise. 
Assume we have an empirical estimate of the noisy frame operator $\tilde{S}$ and it is still invertible, which happens if $\{\tilde{\Pi}_{z,g}\}_{z,g}$ is still informationally complete. 
We can make use of this estimate in the construction of a scalar function $\tilde{o}(z,g) = (O | \tilde{S}^{-1} | \Pi_{z,g})$ that we evaluate instead of $o(z,g)$ in the classical post-processing.
The resulting mean estimator $\hat{\tilde{o}}$ is again unbiased with $\mathbb{E}(\hat{\tilde{o}}) = (O | \tilde{S}^{-1} \, \tilde{S} | \varrho)$. 

When $\nu$ is the uniform measure over the multi-qubit Clifford group, the ideal frame operator $S$ takes the particular simple form of a global depolarizing channel with effective depolarization strength $1/(d+1)$ \cite{Scott2006}.
The simple structure becomes apparent when realizing that $S$ is a channel twirl of the dephasing channel $\mathcal{M} = \sum_{z \in [d]}|\Pi_{z})( \Pi_{z}|$ that describes the computational basis measurement.
For the Clifford group, a unitary $2$-design, $S$ commutes with arbitrary unitary channels \cite{Harrow2013}.
If $\nu$ is the uniform measure on the $n$-fold tensor product of single-qubit Clifford unitaries, $S$ is only invariant under local one-qubit unitary gates. 
Since $\mathcal{M}$ is the tensor-product of local dephasing channels, the projection of $\mathcal{M}$ onto the locally unitarily invariant subspace is the product of local depolarizing channels. 

If the noise does not break the measures' invariance property, $\tilde{S}$ has the same structure as $S$, a global (linear span of local) depolarizing channel(s) for $\nu$ the uniform measure over the global (local) Clifford group. 
This insight is at the heart of the robust shadows introduced in Ref.~\cite{Chen2021}.
There, the invariance assumption is justified by assuming a noise channel $\Lambda$ acting \emph{directly before the measurement} and \emph{independent} of the unitary rotation $g$ that is implemented. 
Under this assumption, $\tilde{S}$ simply is the ideal channel twirl of the concatenation of the dephasing channel and the noise channel. 

More concretely, recall that conjugation with the Clifford group (as with the unitary group) acts irreducibly on the space of traceless matrices, \eg~\cite{Roth2021}. 
Correspondingly, tensor-products of single-qubit Clifford channels decompose into $d$ irreducible representations (\textit{irreps}) without multiplicities, labeled by $k \in \{0, 1\}^n$. 
The projectors onto the irreps are $\Phi_{k} = \bigotimes_{l \in [n]} \Phi_{k_{l}}$ with $\Phi_{0} = d^{-2} \, |\id_{2} )( \id_{2}| $ and $\Phi_{1} = \id_{4} - \Phi_{0}$. 
Assuming single-qubit Clifford-invariance, the noisy frame operator can, thus, be written as
\begin{align}\label{eq:local_invariant_S}
  \tilde{S} = \sum_{k \in \{0,1\}^n} f_{k} \, \Phi_{k} \, 
\end{align}
with coefficients $f_k \in \mathbb R$. 

Ref.~\cite{Chen2021} proposed the following \emph{calibration procedure} to learn the coefficients $f_k$: 
Perform a randomized bases measurement on the input state $\ketbra{0}$. 
The post-processing is analogous to that of linear shadow estimation. 
But here we evaluate the scalar function 
\begin{align}
  \phi_{k}(z,g) = (Z_{k} | \Pi_{z,g}) \, 
  \label{eq:estimator_of_f}
\end{align}
with $Z_k = \bigotimes_{l \in [n]} Z^{k_l}$ and $Z$ the Pauli-$Z$ operator. 
In contrast to the scalar function $o$, $\phi$ does not use the dual frame. 
Assuming the initial state $\ketbra{0}$ can be accurately prepared, the mean of $\phi_{k}$ evaluated on $\mathcal S$ is an unbiased estimator for $f_{k}$. 
From the form of $\tilde{o}(z,g)$ and Eq.~\eqref{eq:estimator_of_f}, we see that the spectrum of $\tilde{S}$ must be characterized to relative precision to guarantee controlled additive precision in estimator $\hat{\tilde{o}}$.

\begin{figure}[t!]
\centering
\includegraphics[width=\columnwidth]{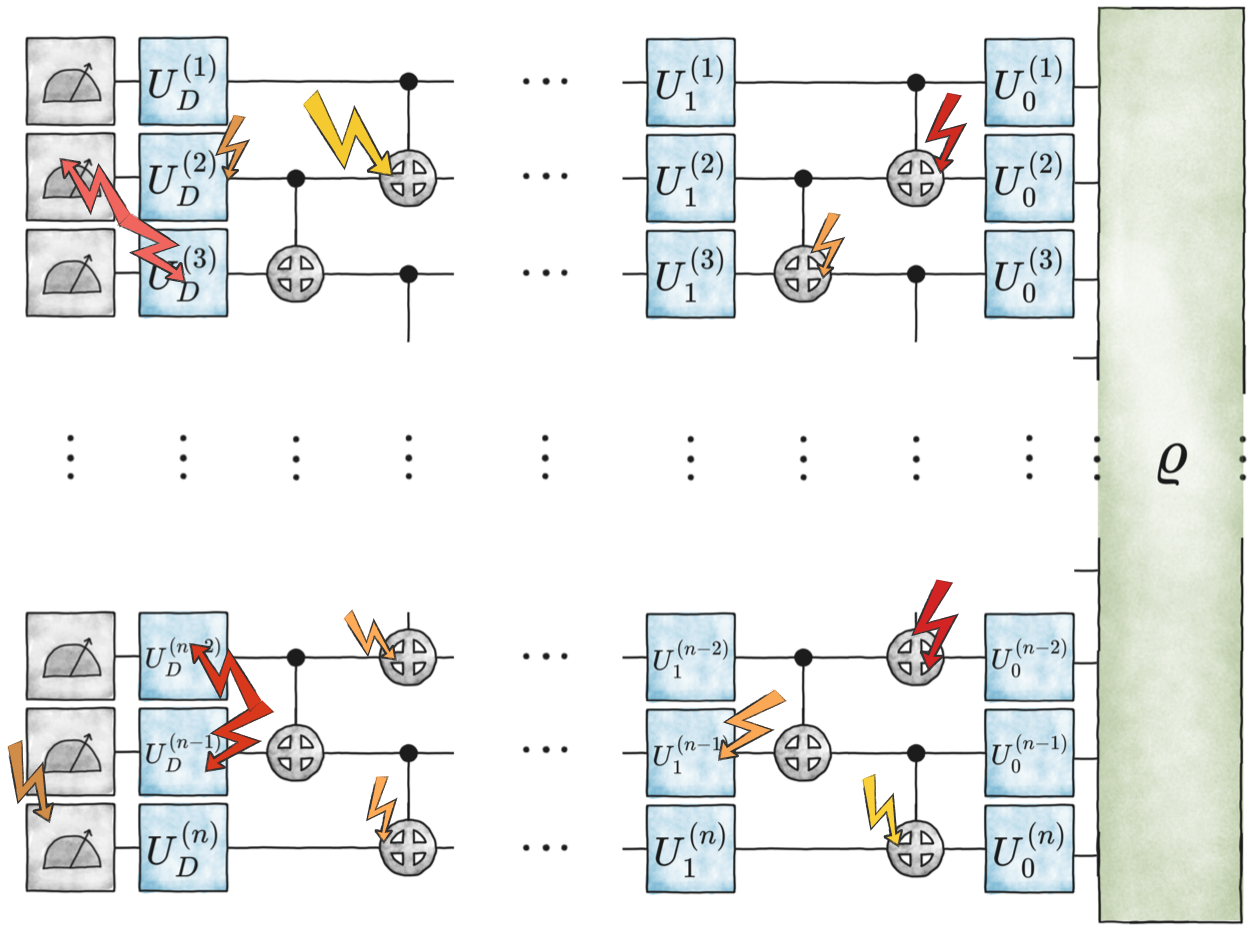}
\caption{
  \textbf{Hardware-efficient circuits for robust shallow shadows.}
  Given an initial state $\varrho$, the circuit implementing the measurement consists of $(D + 1)$ random layers. 
  For $D = 0$, only a layer of $n$ single-qubit rotations is present.
  Each extra layer is composed of a sub-layer of entangling gates followed by $n$ single-qubit rotations. 
  At the end, a computational basis measurement is performed. 
  We assume that the implementation of the circuit can suffer from noise that does not break the local invariance and low-rank representation of the measurement map. For instance, approximately local gate-dependent non-Markovian noise (indicated with the lightning symbols) affecting the gates after the first gate-layer and the read-out is admissible.
}
\label{fig:low-depth-circuit}
\end{figure}

\paragraph{Results.}
\label{sec:results}

The starting observation for our work is the following: 
Let $\nu$ be an arbitrary random ensemble of quantum circuits whose implementation suffers time-stationary, gate-dependent noise and errors. 
The noise can be non-Markovian.
If the first gate layer of the circuits, that is applied to the state under scrutiny, accurately implements local channel twirls, then $\tilde{S}$ in Eq.~\eqref{eq:noisy_frame_definition} still has the structure of Eq.~\eqref{eq:local_invariant_S} even in the presence of noise.
Hence, $\tilde{S}$ can be estimated via the above calibration procedure introduced for local Clifford rotations in Ref.~\cite{Chen2021}. 
A practically motivated assumption is that the major sources of noise in the implemented circuits are entangling gates and readout noise. 
If the first layer then consists of accurately implemented Haar random unitary or Clifford single qubit gates, the invariance assumption holds. 

\paragraph{Circuits and tensor network methods.}
\label{sec:tensor_network_methods}

If the quantum circuit implementing $g$ has low depth, we can make use of tensor network methods at multiple places of the protocol to reduce the runtime and storage complexity of the calibration protocol as well as partially for the estimation protocol.  
To be concrete, we specify a circuit ensemble that is directly motivated by existing hardware architectures and applications. 
The circuit architecture we use is based on the so-called \textit{hardware-efficient ansatz} \cite{Kandala2017}, depicted in Fig.~\ref{fig:low-depth-circuit}. 
The circuit starts with a layer of single-qubit rotations, i.i.d.\ sampled from the Haar measure on single-qubit unitaries or Clifford unitaries. 
This layer is labeled as depth $D = 0$.
Every subsequent layer is composed of a \quotes{sub-layer} of entangling gates followed by another sub-layer of single-qubit rotations sampled from the same measure as before. 
In principle, for the subsequent single qubit layers also other measures are admissible. 

The entangling layer of the hardware-efficient ansatz has a matrix-product operator (MPO) representation of bond dimension {$\chi_{C} = 2$}, see App.~\ref{app:MPO_entangling_layer}.  
Thus, $\Pi_{z,g}$ for $g$ of depth $D$ is exactly described by a matrix-product state (MPS)---also known as tensor train (TT)---of bond dimension (TT-rank) {$\chi_{C}^D$}. 
As a consequence, also the vector of scalar functions $\phi(z,g) = (\phi_k(z,g))_{k \in [n]}$ is described by a TT of rank at most {$\chi_{C}^{2D}$} and can be efficiently computed and stored for every observed tuple $(z,g)$.  

Furthermore, if the noise does not introduce sizable non-local correlation, also $\tilde S$ has an efficient tensor representation, and the coefficient vector $f = (f_k)_{k\in [n]} \in \mathbb R^n$ is given by a low-rank TT.  
We illustrate this argument with a specific noise model in App.~\ref{app:MPO_entangling_layer}.  
In particular, for single-qubit Pauli noise, the TT-rank of $f$ is at most {$\chi = 4^{D}$}.  
This justifies formulating the post-processing of the calibration procedure as the TT recovery problem 
\begin{align}
f = \operatorname*{arg\,min}_{\text{$f'$ rank-$\chi$ TT}} \left\| \, \frac{1}{|\mathcal S|} \sum_{(z, g) \in \mathcal{S}} \phi(z,g) - f^{\prime} \, \right\|_{\ell_2}\,,
\label{eq:TTrecoveryProblem}
\end{align}
\ie the problem of fitting an average of low-rank TTs with a TT of low-rank. 

Standard approaches for tensor recovery such as modified alternating least-square optimization (MALS) \cite{Oseledets2012} can be readily adopted. 
In particular, for MALS, we can perform the local update step for every core using contractions and pseudo-inverses with storage and time complexity scaling polynomially in the system size, TT ranks, and number of samples. 
This suggests that for sufficiently local noise and shallow circuits, $\tilde{S}$ can be efficiently learned.  
Below, we numerically investigate the performance of the scheme using MALS. 
Note, however, that analytically establishing the overall efficiency would additionally require proof of a guarantee for the TT recovery algorithms from $\mathcal{S}$ with polynomial cardinality.

The construction of the dual frame for the estimation protocol uses the inverse of $\tilde S$.  
Since Eq.~\eqref{eq:local_invariant_S} is an orthogonal decomposition, $\tilde S^{-1} = \sum_{k\in\{0,1\}^n} (f_k)^{-1} \Phi_k$. 
From $f$ in TT format, we can efficiently evaluate entries of its element-wise inverse $f^{-1}$ and, thus, the eigenvalues of $\tilde S^{-1}$. 
For certain observables that themselves have an efficient tensor-network representation, having $f^{-1}$ in low-rank TT format would even allow for efficiently computing the estimators $\hat{\tilde o}$.  
Even though $f$ is a low-rank TT, $f^{-1}$ might have a higher rank. 
Finding a low-rank MPS representation for the inverse of a low-rank MPS is, in general, an open question.
Nonetheless, one can attempt to find a low-rank TT description either from the estimate for $f$ or directly from the data using the TT recovery algorithms.

\begin{figure}[t!]
  \centering
  \includegraphics[scale=0.34]{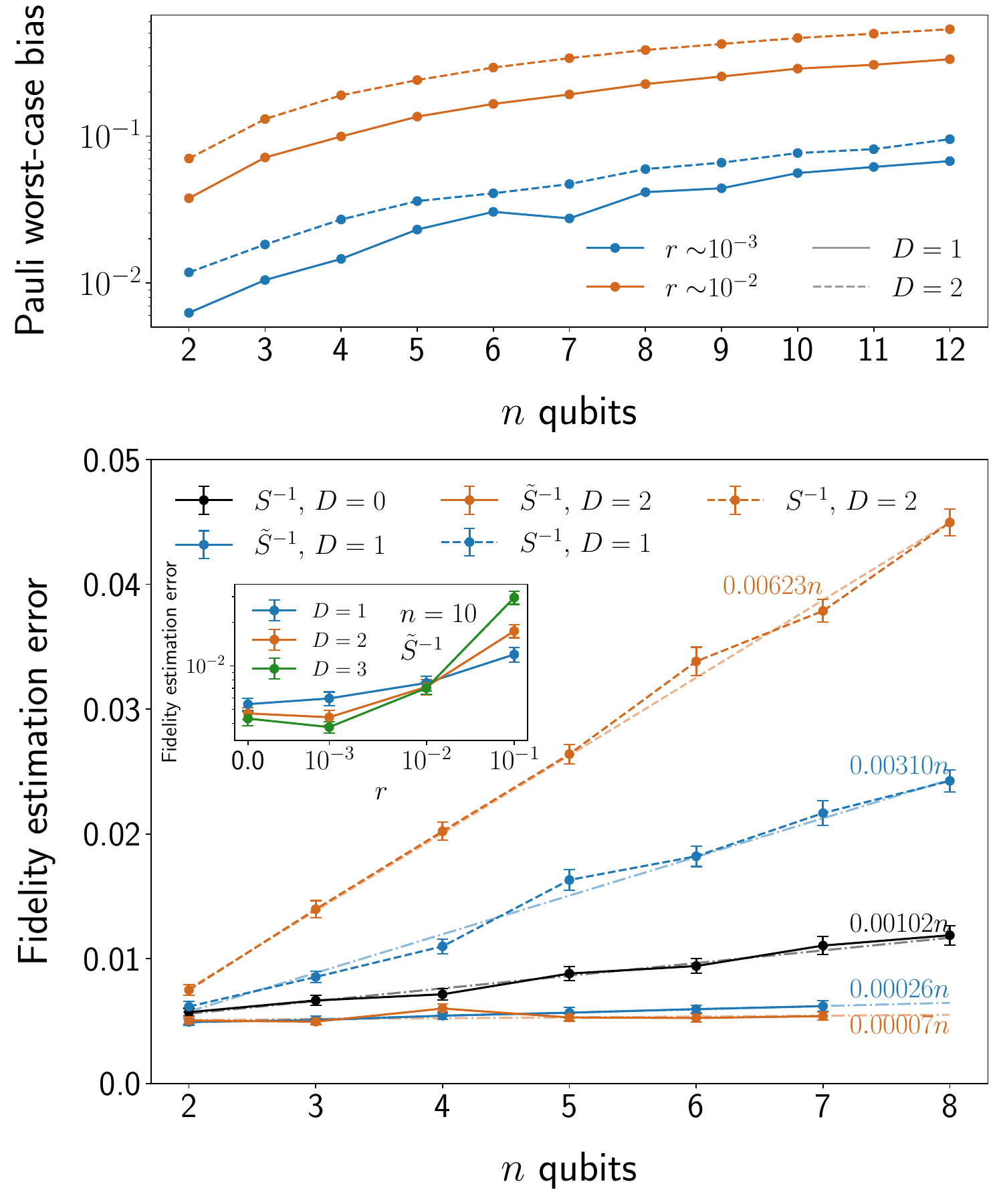}
  \caption{
  \textbf{Numerical experiments under two-qubit incoherent noise model.}
    \textbf{(top)} Worst-case bias for the estimation of Pauli-observables without mitigation for different system sizes, depth, and noise levels. 
    Single-qubit layers sampled from the Haar measure. 
    Noise level indicated by $r$, the average gate infidelity per CNOT, and samples $\abs{\mathcal{S}} = 10^{6}$. 
    We observe a sizable bias about an order of magnitude larger than the two-qubit infidelity and moderately increasing with the system size.
    \textbf{(bottom)} Error in fidelity estimation due to finite statistics $\abs{\mathcal{S}} = 10^5$ for different system sizes, using both $\tilde{S}^{-1}$ and $S^{-1}$. 
    Single-qubit layers sampled from the Haar measure. 
    The error is averaged over $50$ perfectly prepared uniformly random pure states; 
    errors bars denote the standard error. 
    Noise only affects CNOT gates with $r = 10^{-3}$. 
    We observe a reduction of the system-size dependence and, in consequence, smaller estimation errors for shallow circuits compared to (noiseless) local basis measurements ($D = 0$).
    The error in unmitigated fidelity estimation grows linearly with $n$, with a higher slope for increasing depth.
    In contrast, for mitigated fidelity estimation the slope decreases with increasing depth. 
    \textbf{(inset)} Fidelity estimation error using $\tilde{S}^{-1}$ as a function of $r$ for $n = 10$ qubits and $\abs{\mathcal{S}} = 10^5$. 
    We observe a cross-over point at $r = 10^{-2}$; for smaller (larger) infidelities the error decreases (increases) with circuit depth.
  }
  \label{fig:numerical_results_incoherent}
\end{figure}

\paragraph{Numerical results.}
\label{sec:numerical_results}

Here, we numerically study the performance of robust estimation with shallow circuits using architectures that are motivated by applications. 
We use two noise models: 
($i$) two-qubit \emph{incoherent} noise model, where a different unitary noise is applied to a circuit at each execution (see App. \ref{app:incoherent_noise_model} for details);
($ii$) a composition of a two-qubit Pauli noise with \emph{gate-dependent} single-qubit Pauli noise, where different single-qubit gates suffer from different Pauli noise channels.

We begin by assessing the need to mitigate noise in shallow circuits. 
The top panel of Fig.~\ref{fig:numerical_results_incoherent} shows the worst-case bias, $\max_{O, \varrho} \, \operatorname{bias}(O, \varrho)$, 
of standard non-mitigated shadow estimation for Pauli observables (see App.~\ref{app:worst_case_bias} for details) for different system sizes, circuit depths, and incoherent noise levels.
The worst-case bias is monotonically increasing with all three parameters and the results are consistent with a linear increase of the bias with both the circuit depth and system size. 
For two-layer shadows, we observe Pauli worst-case biases that are already for three to four qubits about one order of magnitude larger than the average infidelity of the CNOTs used in the circuit. 
For example, targeting an estimation precision of $10^{-2}$, even fairly accurate CNOTs gates with infidelity $10^{-3}$ can introduce sizable biases in the linear estimation. 
Pauli observables with large non-trivial support tend to have larger biases, see App.~\ref{app:pauli_support_norm}.
We conclude that without mitigation using even very shallow shadows is typically not possible in practice already for small system sizes. 
This underpins the motivation to study our noise-robust variant.

A key application of shadows with global Clifford rotations is fidelity estimation. 
Here, the sample complexity is constant in the system size, while randomized local basis measurements have exponential scaling \cite{Huang2020}.  
This leads to the question of whether an ultra-shallow hardware-efficient ansatz can improve the sample efficiency even in the presence of experimental imperfections. 
In other words: \emph{Can the theoretically established sample-efficiency of classical shadows with entangling bases be translated into a practical advantage?} 
We simulate the protocol of robust shadow fidelity estimation assuming perfectly prepared pure input states. 
The fidelity is estimated w.r.t.\ the input state itself. 
Only the two-qubit gates implementing the basis rotations for the shadow suffer from incoherent noise with average infidelity $10^{-3}$, and the number of samples is $\abs{\mathcal{S}} = 10^5$. 
The noisy frame operator is estimated with $\abs{\mathcal{S}} = 10^6$, which we found sufficient for ensuring that the estimation error of $\tilde{S}$ does not limit the accuracy of the fidelity estimation.  

Fig.~\ref{fig:numerical_results_incoherent} (bottom) shows the estimation error averaged over multiple uniformly random pure states for different system sizes and circuit depths. 
We find that the system-size dependence is significantly reduced already for shadows with $D = 1, 2$ when compared to local circuits. 
In effect, the estimation error is reduced by a factor of about $2$ already for these small systems of size $n = 7, 8$ when using short-depth circuits. 
Importantly, we observe that to harness these effects, the empirical estimation of $\tilde{S}$ put forward here is essential. 
Without noise mitigation, the noise-induced bias, depicted in the figure with dashed lines, deteriorates fidelity estimation by factors of $3$ to $4$ for the same small system sizes and $D = 1$.
For $D = 2$ the estimation error using the noiseless frame operator $S$ is already almost one order of magnitude worse than the one that uses $\tilde{S}$.

\paragraph{Limitations of classical shadows in practice.}
In general, we expect that even assuming perfect bias mitigation, 
the improvements of the estimation error by going to deeper circuits will be limited by incoherent noise.
Intuitively, every CNOT with incoherent errors will increase the probability that one only learns compromised information about the state. 
Even if the noise is perfectly mitigated by accurately estimating $\tilde{S}$, with some probability we learn less about the state. 
Thus, there will be an increased sample complexity of the fidelity estimation itself. 
At the same time, longer circuits (without noise) will improve the variance of fidelity estimation eventually reaching the efficiency of $3$-designs. 
We therefore have two competing effects on the sample complexity. 
We expect that for many estimation tasks, there exists a cross-over point depending on the infidelity of the entangling gates.  
Below this cross-over infidelity, increasing the depth improves estimation accuracy (at a fixed number of samples). 
Above it, accuracy is deteriorated by deeper circuits. 

\begin{figure}[t!]
  \centering
  \includegraphics[scale=0.33]{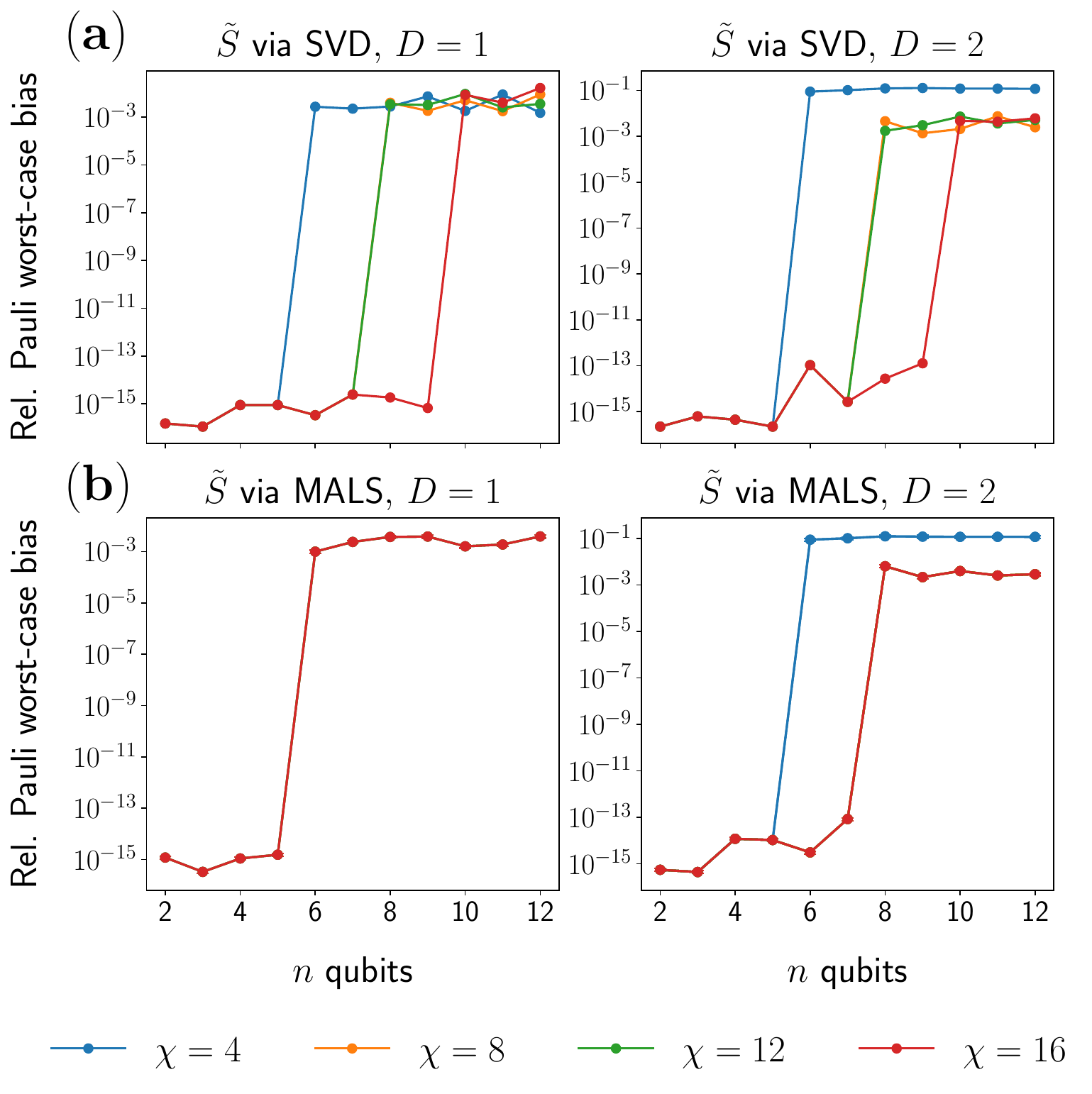}
  \caption{\label{fig:TTrecovery}
    \textbf{Tensor network representation of robust measurement frames of shallow circuits.}
    \textbf{(a)} Worst-case bias for the estimation of Pauli observables introduced by converting $\tilde{S}$ to a matrix product state (TT) representation using sequential SVDs, 
    relative to the size of the corresponding eigenvalue, as a function of $n$, for different bond dimensions $\chi$, and depths $D = 1, 2$. We observe that TT ranks $4$ and $8$ suffice to describe $\tilde S$ up to statistical error.
    \textbf{(b)} Worst-case bias for the estimation of Pauli observables when learning $\tilde{S}$ directly from the calibration data using a modified ALS method. We find a similar performance as for the sequential SVD. Colors indicate the permitted maximal ranks. Actual rank of the result of the rank-adaptive MALS is $\chi=4$ for $D=1$ ($\chi=8$ for $D=2$).
  }
\end{figure}

Using the same fidelity estimation setting as above, we numerically determine the estimation error for different infidelities of the two-qubit gates and circuit depth, 
averaged over random stabilizer states at fixed system size $n=10$ and $\abs{\mathcal{S}} = 10^{5}$.
We display such a cross-over in Fig.~\ref{fig:numerical_results_incoherent} (inset). 
We observe that, for infidelities smaller than $10^{-2}$, the estimation error using $\tilde{S}^{-1}$ is decreased with circuit depth.  
For larger infidelities, deeper circuits suffer from large estimation errors.  
We conclude that indeed even with the robust estimation scheme, deeper shadows being experimentally beneficial is dependent on the noise strength. 

We now take a more detailed look at the TT structure of the frame operator and its inverse. 
We begin with using sequential SVDs to solve the recovery problem in Eq.~\eqref{eq:TTrecoveryProblem} for different ranks $\chi$ using the same noise setting as above. 
Fig.~\ref{fig:TTrecovery} panel (a) shows the additional (relative) worst-case bias in estimating Pauli observables 
when using a TT approximation to $f$ instead of the empirically learned dense vector. 
We indeed find that for $D=1$ a rank $\chi = 4$ approximation yields a small relative error that is constant with the system size for $n \geq 6$ and higher ranks do not yield an improvement. 
Furthermore, we find that for smaller system sizes the MPS is also fitting the statistical fluctuations. 
For $D=2$ we find that $\chi = 8$ accurately approximates $\tilde S$ up to statistical error. 

We then apply the computationally and storage-efficient MALS algorithm to directly learn a TT representation of $\tilde{S}$ from the shadow data and calculate the induced bias in estimating Pauli observables. 
We initialize the algorithm with the analytical result for $S$ for the local Clifford frame. 
We observe an identical performance to the SVD.  
MALS is rank-adaptive but does not increase the rank beyond $\chi = 8$ for $D=2$ even if permitted.

\begin{figure}[t!]
  \centering
  \includegraphics[scale=0.34]{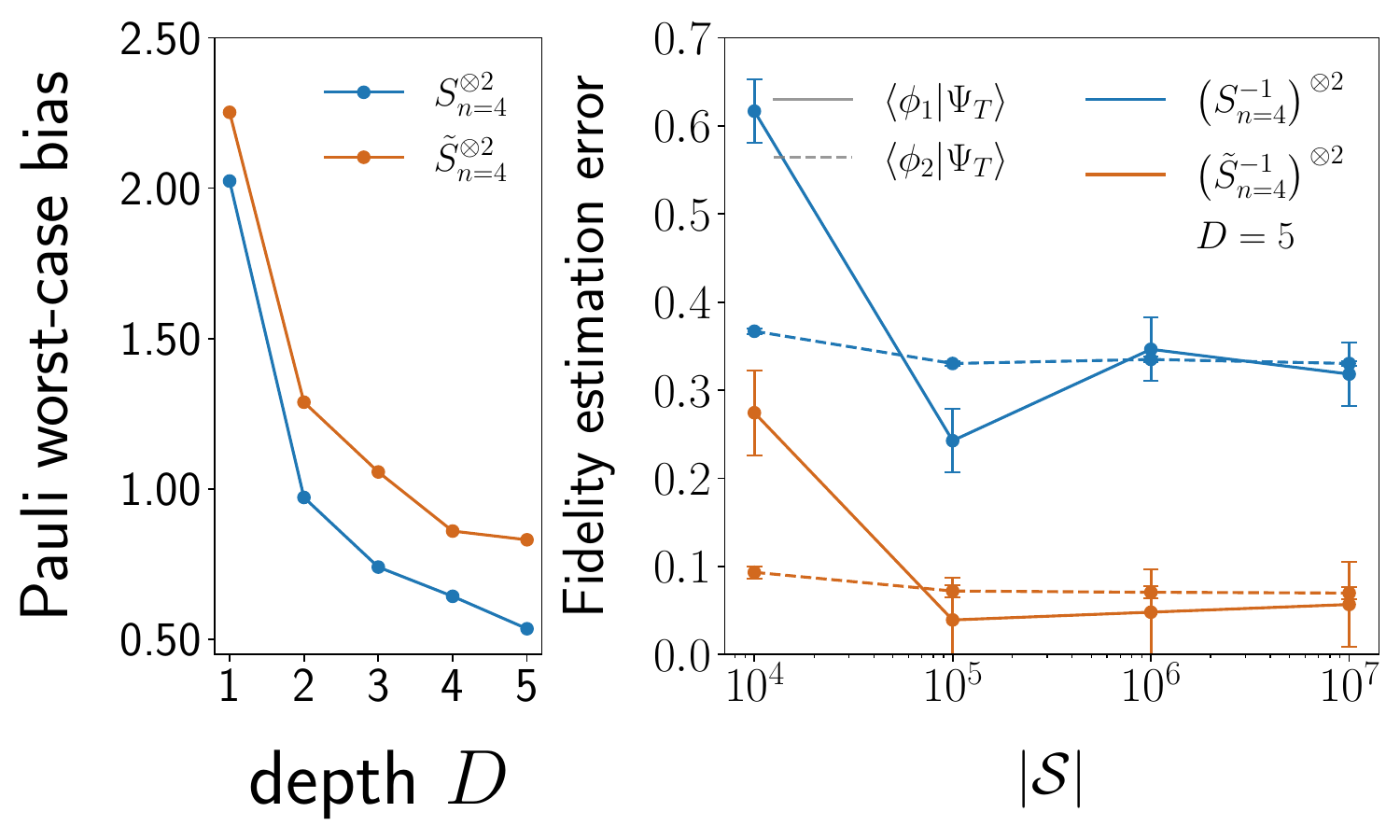}
  \caption{
    \textbf{Simulation of overlap estimation from Ref. \cite{Huggins2022} under a gate-dependent noise model.}
    \textbf{(left)} Worst-case bias for the estimation of Pauli observables introduced by implementing short-depth circuits with and without noise in an experiment, 
    but using the ideal global Clifford frame in the post-processing stage, as a function of $D$,
    for two frames: $S_{n=4}^{\otimes 2}$ (noiseless, blue line) and $\tilde{S}_{n=4}^{\otimes 2}$ (noisy, brown line).
    Frames were estimated with $\abs{\mathcal{S}} = 10^{6}$.
    The noisy frame was generated under a gate-dependent circuit noise as well as readout noise.
    \textbf{(right)} Fidelity estimation error using aforementioned frames. 
    States $\ket{\Psi_T}$, $\ket{\phi_{1}}$, and $\ket{\phi_{2}}$ are defined in Appendix \ref{app:details_fermionic}.
  }
  \label{fig:numerical_results_sycamore}
\end{figure}

\begin{figure*}[t!]
  \centering
  \includegraphics[width=\textwidth]{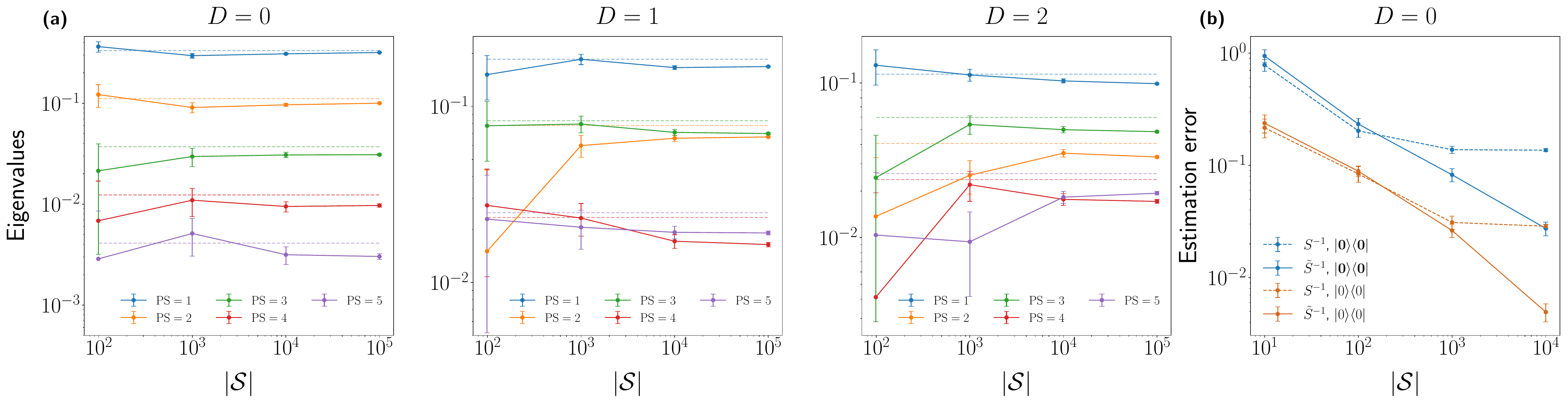}
  \caption{
  \textbf{Experimental realization on \textrm{IBM Quantum}.}
    \textbf{(a)} 
    Eigenvalues of the noisy frame operator $\tilde{S}$ in the Pauli-Liouville basis for $n=5$ qubits, $D \in \{0, 1, 2\}$, and $\abs{\mathcal{S}} = 10^{5}$.
    Error bars correspond to standard errors, \ie to standard deviations divided by $\sqrt{\abs{\mathcal{S}}}$.
    Horizontal dashed lines represent error bars that were masked to avoid displaying negative values in $\log$ scale.
    We plot one eigenvalue per Pauli support $(\mathrm{PS})$.
    Dashed lines show the corresponding eigenvalues if the frames were noiseless.
    \textbf{(b)} Cross-validation for $D = 0$ by using shadow tomography to estimate the fidelity of the state preparation of the single-qubit zero state $\ketbra{0}$ and the $n$-qubit zero state $\ketbra{\mathbf{0}} \equiv \ketbra{0}^{\otimes \, n}$.
  }
 \label{fig:experimental_results}
\end{figure*}

\paragraph{Application.} 
\label{sec:application}

As an application, we simulate the quantum experiment of the optimization protocol that was performed on the Sycamore quantum processor in Ref.~\cite{Huggins2022}.
The authors prepared an $8$-qubit state $\ket{\Psi_{T}}$ and used classical shadows to estimate the wavefunction overlap $\braket{\phi | \Psi_{T}}$, 
where $\ket{\phi}$ belongs to a suitable quantum chemistry basis.
For the specific choices of $\ket{\Psi_{T}}$ and $\ket{\phi}$ in our simulations see App. \ref{app:details_fermionic}.
The measurement circuit architecture used in Ref. \cite{Huggins2022} is the $2$-fold tensor product of $4$-qubit circuits of the same form as Fig.~\ref{fig:low-depth-circuit}.
We call the resulting (noisy) frame operator $S_{n=4}^{\otimes 2}$ ($\tilde{S}_{n=4}^{\otimes 2}$).
Here, the single-qubit rotations are uniformly sampled from the one-local Clifford group.
We simulate the experiment using the \textit{gate-dependent} Pauli noise model.
More specifically, different single-qubit Clifford gates suffer from different Pauli noise channels.
We use the readout error model for the Sycamore chip described in Ref.~\cite{Huang2022}.

The left panel of Fig.~\ref{fig:numerical_results_sycamore} shows, for different circuit depths $D$, 
the worst-case bias using shallow shadows as an approximation of shadow estimation using the $2$-fold tensor product of the $4$-qubit global Clifford group, $S_{\mathrm{gc}, \, n=4}^{\otimes 2}$.
This is motivated by the strategy of Ref. \cite{Huggins2022}.
We see that $\tilde{S}_{n=4}^{\otimes 2}$ displays slower convergence to $S_{\mathrm{gc}, \, n=4}^{\otimes 2}$ compared to $S_{n=4}^{\otimes 2}$.
This suggests that experimentally performing shallow shadows but analytically inverting $S_{\mathrm{gc}, \, n=4}^{\otimes 2}$ for the linear estimation introduces a bias in the estimation.
This bias is more pronounced in the presence of (gate-dependent) noise, stressing the necessity to correctly characterize the noisy measurement frame and use its inverse for the estimations. 
This is presented in the right panel of Fig.~\ref{fig:numerical_results_sycamore}.

\paragraph{Quantum hardware demonstration.} 
\label{sec:hardware_demonstration}

In Fig.~\ref{fig:experimental_results}, we show the results of an experimental implementation of robust shallow shadows on the \texttt{ibm\_nairobi} quantum processor from \textrm{IBM Quantum}.
We implemented the measurement circuit of Fig.~\ref{fig:low-depth-circuit} in $5$ of the $7$ qubits available in the device, for depths $D \in \{0, 1, 2\}$.
In Fig.~\ref{fig:experimental_results} (a) we plot the estimated eigenvalues of $\tilde{S}$ as a function of the cardinality of the shadow $|\mathcal S|$ up to $\abs{\mathcal{S}}=10^5$.
(We also provide histograms for the distribution of samples in Fig.~\ref{fig:numerical_results_sycamore}~(a) in App.~\ref{app:histograms}.)
For simplicity, we show one eigenvalue per Pauli support. 
The dashed lines show the eigenvalues of the corresponding noiseless frame.
In all three cases, the eigenvalues differ from their noiseless counterparts by a statistically significant margin.
This confirms that the measurement frame to be inverted in practice should indeed be $\tilde{S}$ instead of $S$. 
We further observe that, as expected, with increasing depth the eigenvalues get closer. At $D=2$ the noiseless eigenvalues significantly deviate from each other and also the noise-induced deviation differs for different eigenvalues.  
Thus, an approximate construction with the (robust) global shadows is not permissible.

To cross-validate our results, we separate our $10^{5}$ samples in two batches: 
with a first batch of $90\%$ of the snapshots, we estimate $\tilde{S}$; 
with the remaining $10\%$, in turn, we perform a shadow tomography protocol on the  $5$-qubit state $\ketbra{\mathbf{0}}$  as well as on its marginal state $\ketbra{0}$ on one of the qubits.
For each of the two states, we invert the classical shadows using both $\tilde{S}^{-1}$ and $S^{-1}$.
In Fig.~\ref{fig:experimental_results} (b), we plot the fidelity biases introduced by the mitigated and unmitigated estimations, respectively.
For the unmitigated estimation, one can observe that at $\abs{\mathcal{S}}\approx 10^{3}$ the statistical error becomes smaller than the bias (\ie the systematic error due to the `wrong' inverted frame). 
Increasing the number of samples does not improve the accuracy. 
In contrast, for the robust shallow shadows, the accuracy continues to increase steadily as $\abs{\mathcal{S}}$ goes beyond $10^{3}$.

\paragraph{Conclusions.} 
\label{sec:conclusions}

We presented a robust classical shadow scheme for a broad family of noisy, ultra-shallow measurement circuit ensembles.  
Under plausible noise assumptions, the measurement calibration stage admits an efficient tensor-network representation and can be performed
using standard tensor-network tools.
We illustrated the relevance of our method with both numerics and proof-of-principle experiments on an \textrm{IBM Quantum} device. 
Numerically, for relevant parameter regimes, we showed that while unmitigated shallow shadows with noisy circuits become more biased as the depth increases, robust ones become significantly more precise up to a crossing point.  
We thereby established that classical shadows can significantly improve sample efficiency in practice. At the same time, we established that the detrimental effect of incoherent noise on the sample efficiency fundamentally limits capitalizing on the efficiency of classical shadows in practice.
Experimentally, we observed major error reductions in fidelity estimations. 
Our framework does not require analytical calculations of the effective measurement map but replaces it with an efficient direct estimation protocol. 
This offers versatile, practical estimation protocols that realize the efficacy of classical shadow as much as it is possible on actually existing and near-term quantum devices by using the available hardware-efficient noisy shallow circuits. 

\paragraph{Note added.}
\label{sec:note}

During the completion of this paper, another independent work \cite{Hu2024} appeared that also studies robust shallow shadows. 
After the publication of our preprint, Refs.~\cite{Ippoliti2023} and \cite{Schuster2024} corroborated that global shadows can be de-randomized to log-depth. 
As we argue above, however, at current incoherent noise levels these 
constructions are not expected to improve the precision of shadow estimation and one still needs to resort to robust ultra-shallow shadows.

\paragraph{Acknowledgments.} 
\label{sec:acknowledgements}

We thank Stefano Carrazza and Stavros Efthymiou for their insights on numerical simulations. 
We further thank Micha\l\ Oszmaniec for discussions on noise mitigation. 
We thank Jadwiga Wilkens for the availability of the \textit{Quantum Circuit Library} \cite{Wilkens2023}.

\bibliographystyle{apsrev4-2}
\bibliography{references}

\appendix

\section{Details on tensor network description}
\label{app:MPO_entangling_layer}

\begin{figure}[t!]
  \includegraphics[width=\columnwidth]{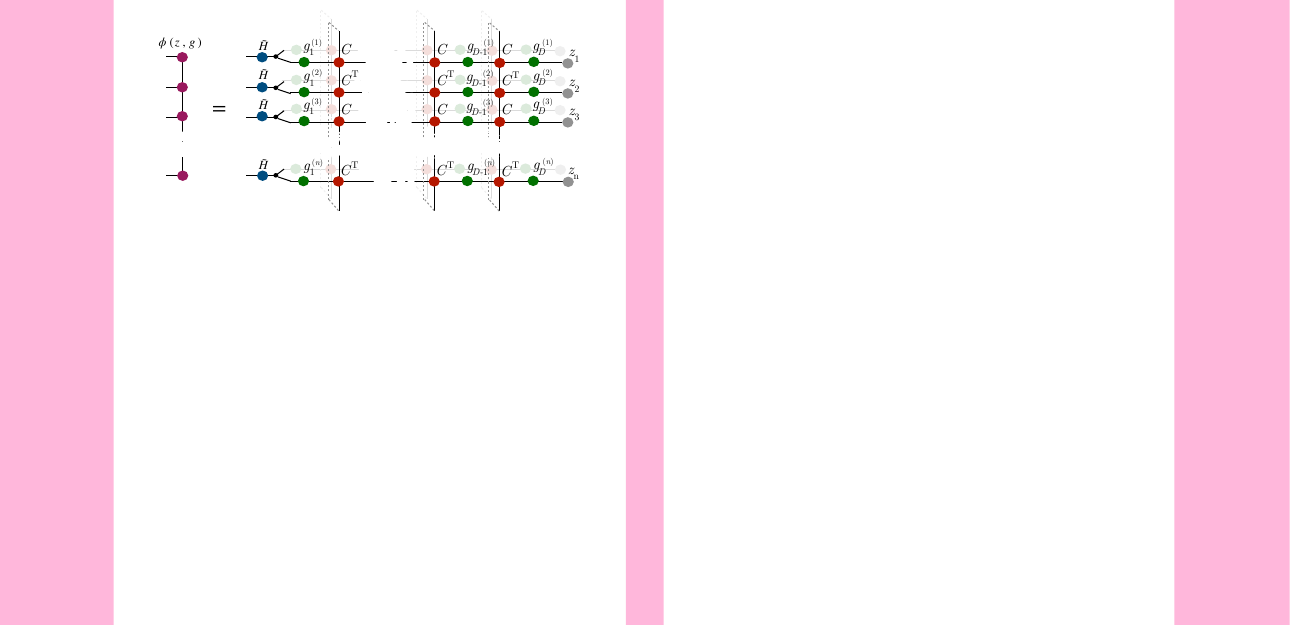}
  \caption{
    \textbf{Tensor network of $\phi(z,g)$}. 
    The network in the background is the conjugate of the one in the foreground determined by the measurement outcome $z$ and the applied random single qubit rotations $g$. 
    The small black dot denotes the copy-tensor, here projecting on the diagonal elements of a density matrix. 
    The matrix $\tilde H$ denotes the Hadamard-Walsh transform. 
    Here it is mapping the diagonal elements of the density matrix to the expectation values of multi-qubit $Z$ observables. 
    The entangling layer with MPO core $C$ and $C^T$ has bond dimension $2$. 
    As a consequence, the TT rank of $\phi(z,g)$ is $4^D$.
  }
  \label{fig:phiTT}
\end{figure}

It is well known that each layer of CNOT gates in the hardware-efficient architecture can be written as an MPO with bond dimension $\chi_{C} = 2$. 
To see this, we decompose the gate as 
\begin{align}\begin{split}
  \operatorname{CNOT} 
  &= \sum_{j \in \{0,1\}} \ketbra{j} \otimes X^{j} \\ 
  &= \left(\sum_{j \in \{0,1\}} \ketbra{j} \otimes \bra{j} \otimes \id \right) \left(\id \otimes \sum_{l\in\{0,1\}} \ket{l} \otimes X^{l}\right) \\
  &= (L \otimes \id) (\id \otimes R)\,,
\end{split}
\end{align}
where $X$ denotes the Pauli-$X$ gate, and the last line defines the operators $L$ and $R$. 
Now we define the MPO core $C_{j,l} \coloneqq (\bra{j} \otimes \id) R \, L (\id \otimes \ket{l}) \in \mathbb C^{2\times 2}$. 
The unitary operator $L_{\operatorname{CNOT}}$ of a $1$D layer of CNOT gates on $n$ qubits with periodic boundary conditions and even $n$ can be written as 
{\small\begin{align}
  \sum_{\substack{j_{1}, j_{2},\ldots \\ l_{1}, l_{2},\ldots}}\hspace{-.2cm}\Tr\left(C_{j_{1}, l_{1}}C_{l_{2}, j_{2}},C_{j_{3}, l_{4}} \cdots C_{j_{n-1}, l_{n}}\right)\!\ketbra {l_1, l_2, \ldots}.
\end{align}}
By construction, $L_{\operatorname{CNOT}}$ is an MPO of bond dimension $2$.
As a consequence $\phi(z,g) = (\phi_k(z,g))_{k\in [n]}$, illustrated in Fig.~\ref{fig:phiTT} is a TT of bond dimension at most $\chi_C^{2D} = 4^D$. 

The operator $\tilde S$ can also be written as a shallow circuit of operators acting on quantum channels. 
Assuming local invariance, also $f = (f_k)_{k\in [n]} \in \mathbb R^n$ has a corresponding circuit representation. 
The rank of the resulting TT description of $f$ is increased by every layer of entangling gates and non-local noise.  
For fairly general noise affecting the shallow circuits, $f$ is of low TT rank.  
To this end, it is instructive to discuss a scenario with strictly local and mostly gate-dependent noise. 
Let us assume that $\Omega^{(l)}_{j}(U)$ is the CPT map describing the noisy implementation of the gate $U \in \mathrm U(2)$ when applied on qubit $l$ in the circuit layer $j$. 
This noise model is general enough to also capture local noise and errors affecting the read-out and entangling gates.  
These noise effects can be included in the respective neighboring layer of single-qubit rotations. 
The local channel twirl with a group $G$ including noise is formally described by the non-commutative Fourier transform of the map $\Omega^{(l)}_{j}$ evaluated at the representation of unitary channels $\omega(U) = U \otimes \bar{U}$, \ie 
\begin{align}
\hat{\Omega}^{(l)}_{j}[\omega] = \int_{G} \bar{U} \otimes U \otimes \Omega^{(l)}_j(U) \mathrm{d}\mu(U)\,. 
\end{align}
We will here simply use $\hat{\Omega}^{(l)}_{j}[\omega]$ as a convenient symbol for the \quotes{noisy channel twirl}. 
Note that $\hat{\Omega}^{(l)}_{j}[\omega]$ is a linear operator on superoperators, a \quotes{super-duper-operator} if one insists.  
We can also introduce a corresponding operator for the entangling layer that conjugates a quantum channel with unitary channels of a layer of CNOT gates. 
The resulting MPO, $\mathcal{C} = \bar{L}_{\operatorname{CNOT}}\otimes L_{\operatorname{CNOT}}\otimes L_{\operatorname{CNOT}}\otimes \bar{L}_{\operatorname{CNOT}}$, has bond dimension $16$.   
The read-out is locally described by $\mathcal{B}_{2} = \sum_{z \in [2]} \rketbra{\Pi_z}$, the local dephasing channel in computational basis. 
Finally, to ensure the local invariance of $\tilde{S}$, we assume gate-independent left noise $\Omega^{(1)}_{j}(U) = \Lambda_{j} \omega(U)$ for the first layer acting on the state under consideration, with $\Lambda_{j}$ being a quantum channel. 

\begin{figure}
  \includegraphics[width=\columnwidth]{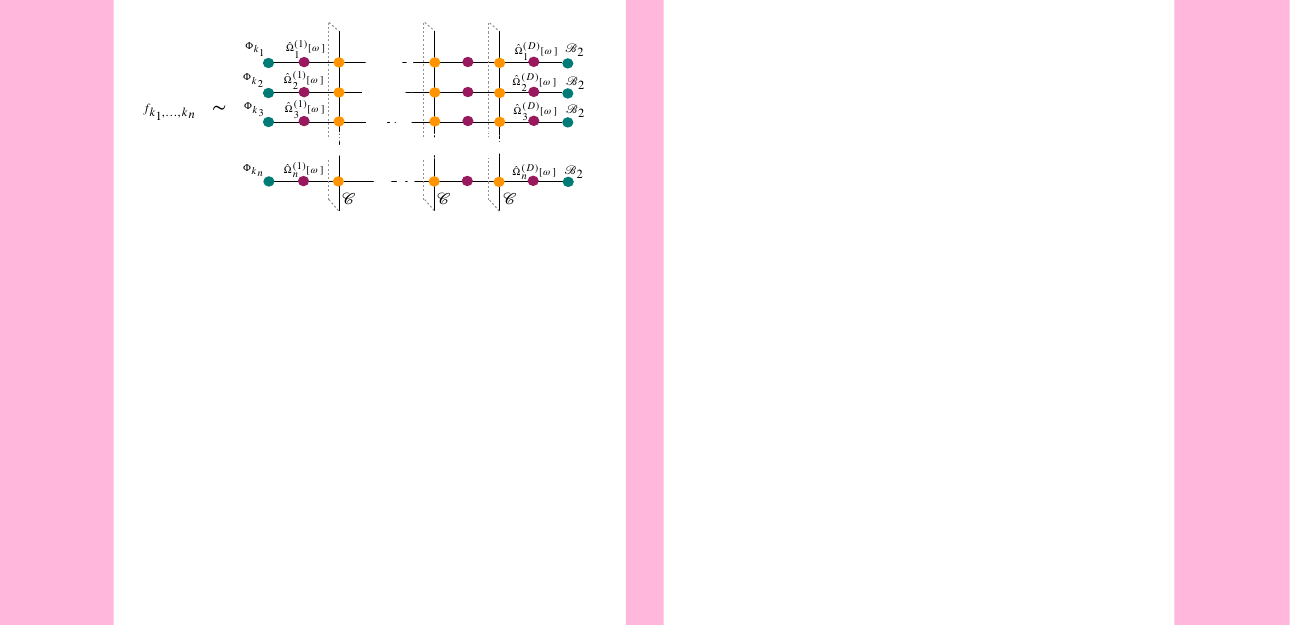}
  \caption{
    \textbf{Tensor network of $f$ for strictly local noise mostly gate-dependent noise.}
    The entangling layer acting here as a \quotes{super-duper operator} on quantum channels has an MPO bond dimension $16$, and restricted to Pauli noise channels it has an MPO bond dimension $4$.
    Correspondingly, the TT rank of $f$ is $16^D$ for local noise and $4^D$ for local Pauli noise. 
   }
   \label{fig:fTT}
\end{figure}

Figure~\ref{fig:fTT} depicts the resulting expression for $f$.  
Without any further assumption, we conclude that $f$ has a TT rank of at most $16^{D}$.  
The rank is even smaller if all noise channels are Pauli noise. 
Then, we can fully restrict the circuit in Fig.~\ref{fig:fTT} as acting only on the subspace of Pauli noise channels, \ie\ diagonal channels in Pauli operator basis. To see this note that the local dephasing channel is a Pauli noise channel. 
Under the assumption of Pauli noise, the noisy channel twirl maps Pauli noise channels to Pauli noise channels. 
Being a Clifford unitary, the entangling layer acts as a two-local permutation on the diagonal of the Pauli-noise channel.  
The representation of $\mathcal{C}$ restricted to Pauli-noise channels has (MPO) bond dimension $4$. 
Hence, $f$ is described by a TT of rank at most $\chi = 4^{D}$.

The argument for one-local noise can be generalized to quasi-local noise by introducing, \eg, MPO descriptions of the implementation map modeling cross-talk effects, or modeling a noisy entangling layer $\mathcal{C}$ with a potentially larger bond dimension.  
In general, the TT rank of $f$ will be increased due to correlated noise effects.  
If these correlations are sufficiently short-ranged, $f$ still has an effective low-rank description. 
We expect that this will often be the case on actual quantum hardware. 

\section{Incoherent noise model}
\label{app:incoherent_noise_model}

We model the incoherent noise as an unitary channel $\Uplambda \coloneqq \exp\left( i \, \gamma \, H_{\text{g}} / 8 \right)$,
where $H_{\text{g}}$ is a random Hamiltonian sampled from the \textit{Gaussian Unitary Ensemble} (GUE) \cite{Fyodorov2004, Brieger2023},
$\gamma$ is the so-called \textit{strength parameter}, and the constant $8$ is a normalization factor (see Sect. IV-B of Ref. \cite{Brieger2023}).
By sampling $H \sim \text{GUE}$, we guarantee that, on average, that is no preferential direction in which the noise acts.
Even though $\gamma \in [0, \, 8]$, we focus on $\gamma < 2$ based on reported error rates for the CNOT gate across multiple commercial quantum computing platforms.

\section{Worst-case bias measure}
\label{app:worst_case_bias}
Consider the case where the half-sided noisy frame operator is still diagonal, \ie $\tilde{S} = \sum_a \tilde{S}_{a} \rketbra{w_{a}}$ with $w_{a}$, $a \in \mathbb F_2^{2n}$, Frobenius-normalized Pauli operators such that $\|w_a\|_F =1$.  
We define the bias of standard shadow estimation of an observable from noisy shadow measurements in state $\varrho$ as 
\begin{align}
  \operatorname{bias}(O,\,\varrho) \coloneqq |\rbraket{O}{\varrho} - \rbra{O} S^{-1}\tilde{S} \rket{\varrho}| \, . 
\end{align}
The norm difference $\| S^{-1} \tilde{S} - I \|_\infty$ has the interpretation of the worst-case bias for estimating a single Pauli observable $O_a = \sqrt{d} \, w_{a}$. 
This can be directly seen from the following expression
\begin{align}
\begin{split} 
  \max_{\varrho, a \in \mathbb{F}^{2n}_2} \operatorname{bias}(O_{a},\varrho)
    &= \sqrt{d} \, \max_{a, \varrho} |\rbra{w_{a}} I - S^{-1}\tilde{S} \rket{w_{a}}\rbraket{w_{a}}{\varrho} | \\
    &\leq \|I - S^{-1}\tilde{S}\|_\infty\,,
\end{split}
\end{align}
where we used the diagonality of $\tilde{S}$ in Pauli basis and that $|\rbraket{w_a}{\varrho}| \leq \|w_a\|_{\infty} \|\varrho\|_1 \leq 1/\sqrt{d}$ for all $a, \varrho$, and $\|\cdot\|_{1}$ denoting the trace-norm (nuclear / Schatten-$1$ norm). 
Most importantly, this bound is saturated for $\varrho$ in an eigenstate of $w_{a}$ the observable suffering the maximal bias. 

\section{Worst-case bias per Pauli support}
\label{app:pauli_support_norm}

\begin{figure}[t!]
  \centering
  \includegraphics[scale=0.28]{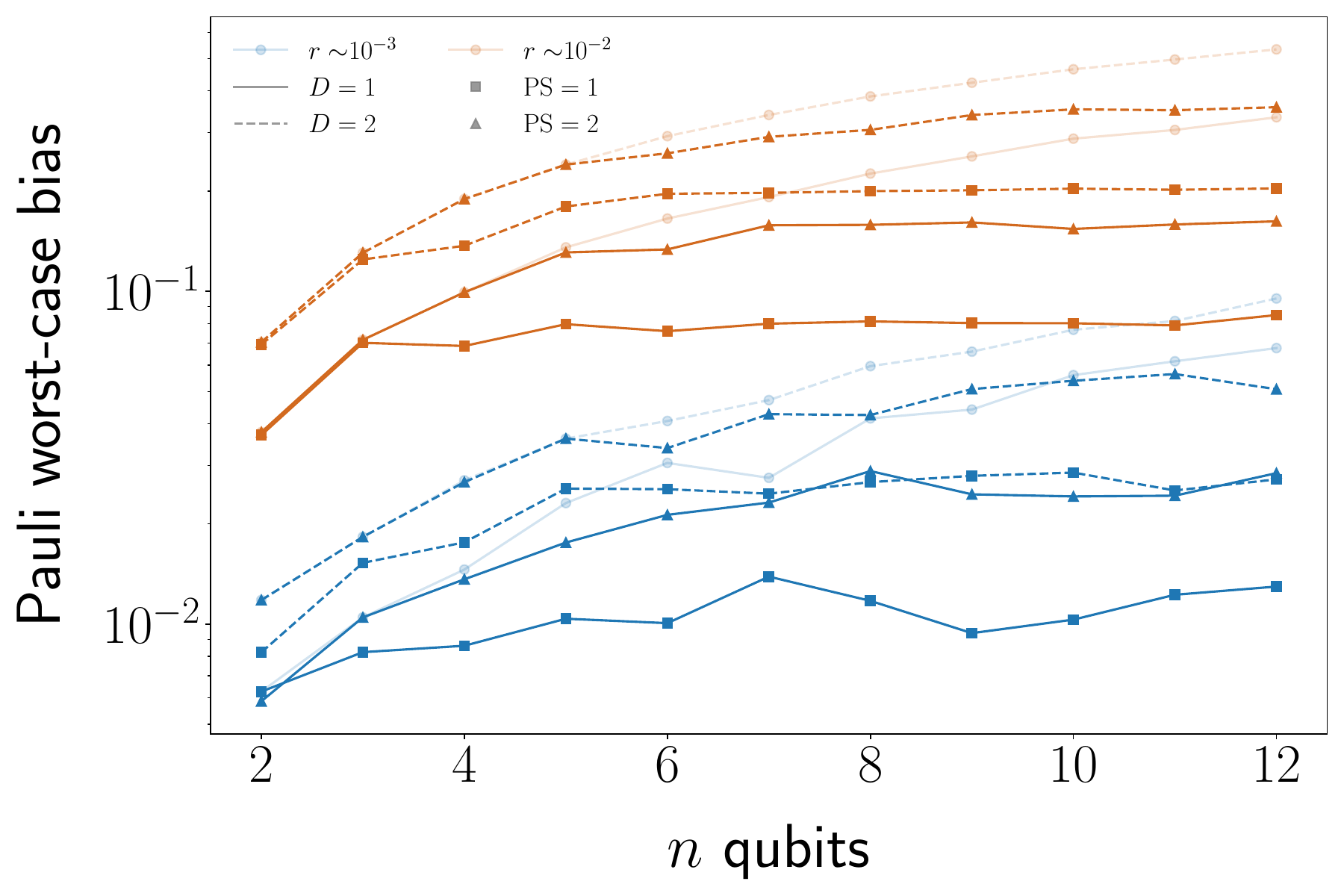}
  \caption{
	\textbf{Worst-case bias from Fig.~\ref{fig:numerical_results_incoherent} for different Pauli supports.}
  }
 \label{fig:pauli_support_norm}
\end{figure}

In Figure \ref{fig:pauli_support_norm} we show the same worst-case bias of standard non-mitigated shadow estimation for Pauli observables, under the same noise conditions.
This is presented in shaded lines.
However, this time we also present the worst-case bias for fixed Pauli supports $\mathrm{PS} = 1, 2$.
We observe that the Pauli worst-case biases for these Pauli supports seem to hit a plateau for both noise levels. 
However, for $r \sim 10^{-2}$, even though stable, the biases are still sizable.

\section{Overlap estimation based on Ref. \cite{Huggins2022}}
\label{app:details_fermionic}

The optimization protocol for quantum chemistry problems implemented by Ref. \cite{Huggins2022} involves estimating the wavefunction overlaps $\braket{\phi | \Psi_{T}}$.
This is done by preparing $\ket{\Psi_{T}}$ in a quantum processor, constructing its classical shadow representation, and evaluating the overlap with multiple $\ket{\phi}$ via post-processing.
The state $\ket{\Psi_{T}}$ is called a \textit{trial function}, and $\ket{\phi}$ is named a \textit{walker function}.
We refer to Ref.~\cite{Huggins2022} for more details about $\ket{\Psi_{T}}$ and $\ket{\phi}$ in the quantum chemistry context.
For our intents and purposes, we set a phase $\theta = \pi / \sqrt{2}$, and
\begin{align}
\ket{\Psi_{T}} &= \cos^{2}(\theta) \ket{11001100} + \frac{1}{2} \sin(2\theta) \ket{11000011} \nonumber \\
& \quad + \frac{1}{2} \sin(2\theta) \ket{00111100} + \sin^{2}(\theta) \ket{00110011} \, .
\end{align}
Moreover, we define 
\begin{align}
\ket{\phi_{1}} &= \frac{1}{2} \ket{11001100} + \frac{1}{2} \ket{11000011} \nonumber \\
& \quad + \frac{1}{2} \ket{00111100} + \frac{1}{2} \ket{00110011} \, .
\end{align}
and
\begin{align}
\ket{\phi_{2}} &= \frac{1}{\sqrt{2}} \ket{11001100} + \frac{1}{\sqrt{6}} \ket{11000011} \nonumber \\
& \quad + \frac{1}{\sqrt{6}} \ket{00111100} + \frac{1}{\sqrt{6}} \ket{00110011} \, .
\end{align}
We simulate the perfect preparation of $\ket{\Psi_{T}}$ and its reconstruction via classical shadows with noisy measurements.
The noise model chosen is based on the one from the Sycamore processor \cite{Arute2019, Huang2022}, 
since Ref. \cite{Huggins2022} implements this protocol in that quantum hardware.

\section{Data histograms for Fig.~\ref{fig:experimental_results} (a)}
\label{app:histograms}

As explained in the main text, given the sequence $\mathcal{S}$ of snapshots obtained using the quantum hardware, the $k$-th eigenvalue of the noisy frame operator $\tilde{S}$ is estimated by calculating $\mathbb{E}_{(z,g) \in \mathcal{S}} \,\, \phi_{k}(z,g)$, $\forall \, k$, with $\phi_{k}$ being defined in Eq.~\eqref{eq:estimator_of_f}.

Here, we present the distributions of the data used to estimate the eigenvalues of the noisy frame operator obtained on the \texttt{ibm\_nairobi} quantum processor using the \textrm{IBM Quantum} platform.
As in Fig.~\ref{fig:experimental_results} (a), we show one eigenvalue per Pauli support $\operatorname{PS}$ and per depth $D \in \{0, 1, 2\}$.
Moreover, we use a sequence of size $\abs{\mathcal{S}} = 10^{5}$.

We observe that multiple eigenvalues significantly deviate from the value for global rotations still for $D=2$.  Furthermore, the bias significantly differs for different eigenvalues.  
Thus, mitigating the bias by rescaling the estimator with a single \quotes{global} scalar or equivalently only relying on ratio estimators does not remove the bias.

\begin{figure*}[t!]
  \centering
  \includegraphics[scale=0.2]{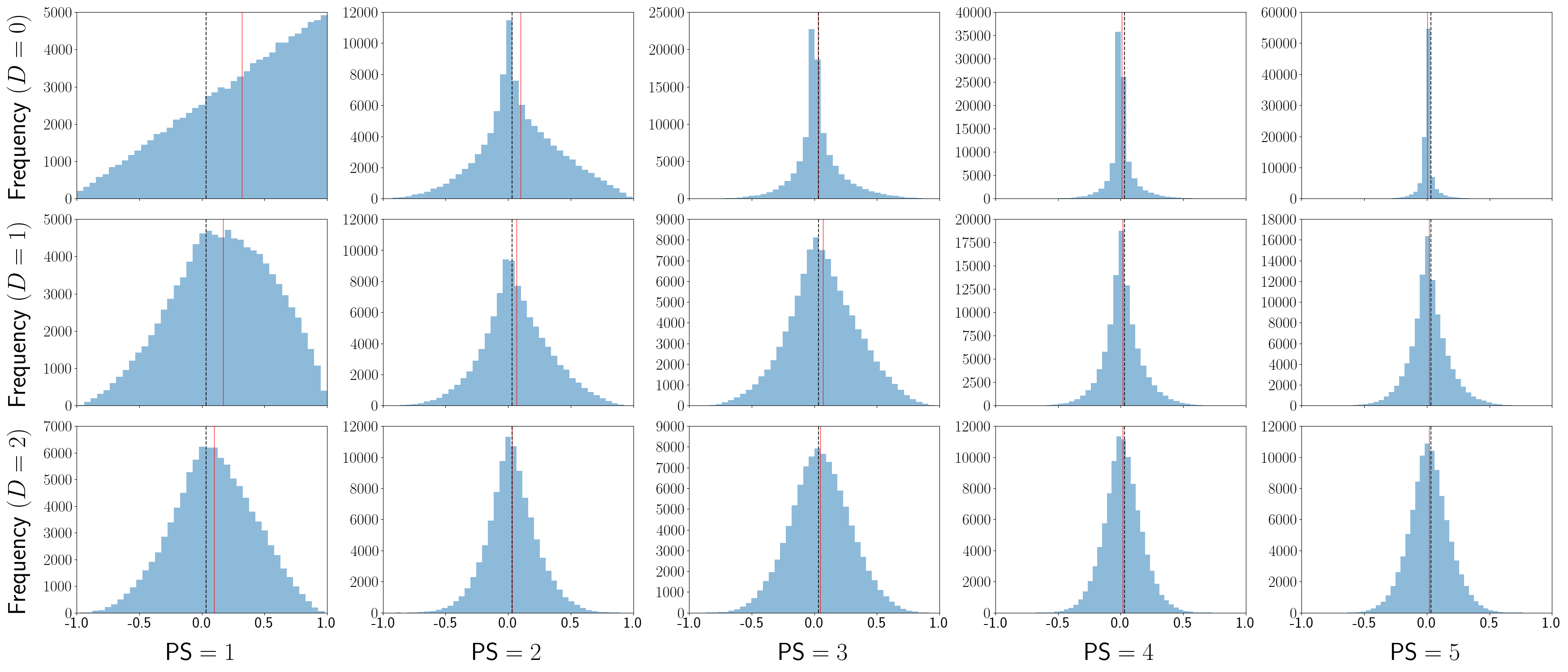}
  \caption{
    \textbf{Histograms of experimental realization on \textrm{IBM Quantum}.}
    Distribution of samples used to estimate eigenvalues of the noisy frame $\mathcal{S}$ for different Pauli supports $1\leq\operatorname{PS}\leq 5$, 
    over $|\mathcal{S}|=10^5$ samples from the \texttt{ibm\_nairobi} quantum processor, for $n = 5$ and $D \in \{0, 1, 2\}$. 
    The average over the $10^5$ samples gives our estimates of the eigenvalues of $\tilde{S}$ shown in Fig.~\ref{fig:experimental_results} (a). 
    We indicate the corresponding eigenvalue by dashed red lines, while the dashed black lines indicate the $1/(d + 1)$ level.
  }
 \label{fig:histograms}
\end{figure*}

\end{document}